# A Step in the Direction of Resolving the Paradox of Perdew-Zunger Self-interaction Correction. II. Gauge Consistency of the Energy Density at Three Levels of Approximation

Puskar Bhattarai[1, #], Kamal Wagle[1], Chandra Shahi[1, 2], Yoh Yamamoto[2], Selim Romero[2], Biswajit Santra[1], Rajendra R. Zope[2], Juan E. Peralta[3], Koblar A. Jackson[3], and John P. Perdew[1, 4]

[1]Department of Physics, Temple University, Philadelphia, PA 19122 USA

[2]Department of Physics, University of Texas at El Paso, El Paso, TX 79968 USA

[3]Department of Physics and Science of Advanced Materials, Central Michigan University, Mount Pleasant, MI 48859 USA

[4]Department of Chemistry, Temple University, Philadelphia, PA 19122 USA

[#]Corresponding author (puskar.bhattarai@temple.edu)

**Abstract**

The Perdew-Zunger(PZ) self-interaction correction (SIC) was designed to correct the one-electron limit of any approximate density functional for the exchange-correlation (xc) energy, while yielding no correction to the exact functional. Unfortunately, it spoils the slowly-varying-in-space limits of the uncorrected approximate functionals, where those functionals are right by construction. The right limits can be restored by locally scaling down the energy density of the PZ SIC in many-electron regions, but then a spurious correction to the *exact* functional would be found unless the self-Hartree and exact self-xc terms of the PZ SIC energy density were expressed in the same gauge. Only the local density approximation satisfies the same-gauge condition for the energy density, which explains why the recent local-scaling SIC (LSIC) is found here to work excellently for atoms and molecules only with this basic approximation, and not with the more advanced generalized gradient approximations (GGAs) and meta-GGAs, which lose the Hartree gauge via simplifying integrations by parts. The transformation of energy density that achieves the Hartree gauge for the exact xc functional can also be applied to approximate functionals. Doing so leads to a simple scaled-down self-interaction (sdSIC) correction that is typically much more accurate than PZ SIC in tests for many molecular properties (including equilibrium bond lengths). The present work shows unambiguously that the largest errors of PZ SIC applied to standard functionals at three levels of approximation can be removed by restoring their correct slowly-varying-density limits. It also confirms the relevance of these limits to atoms and molecules.

# 1. Introduction: the failure of Perdew-Zunger self-interaction corrections in the slowly-varying density limit

Density functional theory (DFT) [1,2] is a practical approach to the many-electron problem for the groundstate energy and density, requiring only the solution of selfconsistent one-electron Schrödinger equations. Starting from a formally-exact DFT, approximations are made for the expectation value of the electron-electron interaction plus the correlation contribution to the kinetic energy, replacing this sum by the exactly-treated classical Coulomb or Hartree energy $U[n]$ that depends nonlocally upon the electron density $n(\vec{r})$ plus an approximate exchange-correlation energy $E_{xc}^{app}[n_\uparrow, n_\downarrow]$, a functional of the spin-up and spin-down electron densities. The approximation is typically a *single* integral over three-dimensional space of a modelled energy density that depends upon the local spin-up and spin-down electron densities, and sometimes on their gradients or other position-dependent ingredients. That makes a DFT calculation much more computationally efficient than calculations based on the many-electron wave function. Various models for the energy density have been developed over the years, yielding better predictions for the properties of atoms, molecules, and solids by matching more of the conditions known to be satisfied by the exact but unknown functional, $E_{xc}^{exact}$. For example, the strongly constrained and appropriately normed SCAN functional [3] satisfies all 17 of these conditions that a meta-generalized gradient approximation can satisfy, and is remarkably accurate for the description of the properties of molecules and condensed matter [4], including such complex systems as liquid water [5] and the solid cuprates [6]. SCAN and its simpler nonempirical predecessors, going back to the spin-density version of the local density approximation (LDA) [1,7], are exact by construction for any system of uniform electron density. Further refinements in the description of the slowly-varying limit are added along the path from LDA to SCAN.

One condition of the exact $E_{xc}$ that is violated by many density functional approximations (DFAs) is one-electron self-interaction freedom. For a one-electron density $n_{\alpha\sigma} = |\varphi_{\alpha\sigma}|^2$, where α and σ are the orbital and spin indices respectively,

$$E_{xc}^{exact}[n_{\alpha\sigma}, 0] = -U[n_{\alpha\sigma}] \tag{1}$$

for the exact functional, but this equality is lost for many approximate functionals, resulting in a residual self-interaction of the electrons in DFA calculations. The exact equality holds even when $\varphi_{\alpha\sigma}$ is noded [8]. Self-interaction errors (SIE) manifest in various ways in DFT calculations. Two prominent examples are 1) that orbital energies for the highest occupied electron states are too high compared to minus the corresponding electron removal energies and 2) that barrier heights for chemical reactions are predicted to be too low by DFT in comparison to accurate values taken from high-level reference calculations. Perdew and Zunger [9] introduced a scheme to make approximate functionals one-electron self-interaction free. The Perdew-Zunger self-interaction correction (PZ-SIC) removes self-interaction on an orbital-by-orbital basis. It has the important features that it is exact for all isolated one-electron densities and would yield no correction if applied to the exact functional. Applying PZ SIC improves the performance of DFAs in situations that are dominated by self-interaction error (SIE), including stretched radicals (e.g., [10]), but, paradoxically, it does so at the expense of accuracy in situations where SIE is not significant [11]. For example, equilibrium properties such as molecular atomization energies are predicted quite accurately by SCAN, but only poorly by SCAN-SIC [8,12]. Even in situations where PZ SIC improves predictions, such as atomic polarizabilities [13] or molecular dipole moments[14], there is often an overcorrection in which an underestimation by the uncorrected functional becomes an overestimation by its PZ-SIC version or *vice-versa*.

It was recently shown [15] that PZ SIC disrupts the behavior of nonempirical density functional approximations (DFAs) such as the local density approximation (LDA) [1,7], the Perdew, Burke and Ernzerhof (PBE) [16] generalized gradient approximation (GGA), and the SCAN meta-GGA in the slowly-varying density limit, introducing non-negligible errors of a few percent of the exchange-correlation energies for uniform electron densities where the uncorrected approximations are exact.

In this paper we analyze an alternative to PZ SIC that addresses this shortcoming by formally scaling down the correction terms in regions where the density is slowly varying, reducing it to zero for any uniform density while leaving it intact in regions where the density has single orbital character. An "exterior" scaling by an orbital-dependent constant for the SIC contribution to the energy from each occupied orbital (depending upon the orbital and upon the degree of orbital overlap in the region where the orbital is located) was proposed earlier [17,18] to guarantee the correct uniform-density limit. Interestingly, we show here that, by making a gauge transformation that renders the self-Hartree and self-exchange-correlation energy densities gauge-consistent for the exact functional, what would otherwise be an interior scaling of the PZ-SIC correction terms at each point in space becomes an exterior scaling for each orbital contribution. We have tested this scaled-down self-interaction correction or sdSIC approach for a variety of properties of atoms and molecules, and for three levels of the uncorrected approximation (LDA, GGA, and SCAN meta-GGA). Generally, we find that sdSIC predictions of equilibrium properties are improved relative to the results of PZ-SIC calculations, while the improvement of the latter method in situations where SIE is dominant is largely retained. These results demonstrate that the largest errors of PZ SIC applied to traditional density functionals can be reduced by restoring the correct slowly-varying limit of the uncorrected functional. Since our numerical tests are for atoms and molecules, we thus also confirm the relevance [19] of the slowly-varying limit to chemical systems.

Recently an interior scaling-down of LDA-PZ-SIC (local-scaling SIC or LSIC [20]) was found to achieve remarkable and almost universal improvement over LDA. An important conclusion of the present work is that interior or local scaling works excellently with LDA because the LDA exchange-correlation energy density is in the same gauge as the Hartree energy density, and less well with PBE and much less well with SCAN because their energy densities are not in the same gauge as the Hartree energy density.

## 2. A gauge-consistent, scaled-down self-interaction correction at three levels of the uncorrected functional

The additive orbital-by-orbital PZ SIC to $E_{xc}^{app}[n_\uparrow, n_\downarrow]$ is

$$\Delta E_{xc}^{\alpha\sigma} = -\{U[n_{\alpha\sigma}] + E_{xc}^{app}[n_{\alpha\sigma}, 0]\}, \tag{2}$$

where $\sigma = \uparrow, \downarrow$. Here $n_\sigma = \sum_{\alpha\sigma} n_{\alpha\sigma}$, and the localized occupied orbital densities $n_{\alpha\sigma} = |\varphi_{\alpha\sigma}|^2$ are found from a set of localized (for the sake of size consistency) SIC orbitals $\varphi_{\alpha\sigma}$ constructed as described in section 3 via unitary transformation of the occupied canonical orbitals. There is typically much cancellation between the two terms of $\Delta E_{xc}^{\alpha\sigma}$:

$$U[n_{\alpha\sigma}] = \frac{1}{2}\int d^3r\, n_{\alpha\sigma}(\vec{r}) u([n_{\alpha\sigma}]; \vec{r}), \tag{3}$$

where

$$u([n_{\alpha\sigma}]; \vec{r}) = \int d^3r'\, \frac{n_{\alpha\sigma}(\vec{r}')}{|\vec{r} - \vec{r}'|}, \tag{4}$$

and

$$E_{xc}^{app}[n_{\alpha\sigma}, 0] = \int d^3r\, n_{\alpha\sigma}(\vec{r}) \varepsilon_{xc}^{app}([n_{\alpha\sigma}, 0]; \vec{r}), \tag{5}$$

but the cancellation is incomplete, necessitating the correction.

PZ SIC makes any approximation exact for any one-electron density and gives no correction to the exact functional. But it overcorrects, especially for PBE and SCAN, in many-electron regions of real systems. The iso-orbital indicator $z_\sigma(\vec{r}) = \frac{\tau_\sigma^W(\vec{r})}{\tau_\sigma(\vec{r})}$ distinguishes many-electron from single-electron regions of the total density $n_\sigma = \sum_\alpha |\psi_{\alpha\sigma}|^2$ of spin $\sigma(\uparrow, \downarrow)$. Here

$$\tau_\sigma^W(\vec{r}) = \frac{|\nabla n_\sigma|^2}{8 n_\sigma} \tag{6}$$

and

$$\tau_\sigma(\vec{r}) = \sum_\alpha^{occ} \frac{1}{2} |\nabla \psi_{\alpha\sigma}|^2. \tag{7}$$

This typical meta-GGA ingredient is bounded in the range $0 \leq z_\sigma \leq 1$. Clearly $z_\sigma = 0$ for a uniform density (a many-electron-like region) and 1 for an iso-orbital (one-electron-like) density. In the slowly-varying-density limit, in which $\tau_\sigma$ approaches a Thomas-Fermi limit $n_\sigma^{5/3}$, $z_\sigma$ is of order $\nabla^2$.

In interior scaling, we introduce a scale-down factor

$$f_m(z_\sigma) = m z_\sigma^m - (m-1) z_\sigma^{m+1} \tag{8}$$

to be applied to the energy densities inside $\Delta E_{xc}^{\alpha\sigma}$. The integer $m$ is chosen to keep the slowly-varying limit of the self-interaction corrected exchange-correlation approximation correct through the same order in $\nabla$ as the parent functional. For LDA, PBE, and SCAN, we choose $m = 1, 2,$ and 3, keeping the approximations correct through order $\nabla^0, \nabla^2$, and $\nabla^4$ respectively (although PBE is not fully correct to second order). The $z_\sigma^{m+1}$ term in the scale-down factor ensures that, for each $m$, the scale-down factor approaches 1 just as $f_1(z_\sigma) = z_\sigma$ does (Fig. 1). $f_1(z_\sigma)$ was used as a scale-down factor for LDA in Ref. [20]. As $m$ increases from 1, $f_m(z_\sigma)$ increasingly scales down the SIC in slowly-varying regions *only*. This makes it possible for us to reach firm conclusions about the importance of the slowly-varying limits for atoms and molecules.

Total energies are measurable and unique, and individual contributions thereto are unique, but energy densities are not unique. In the Hartree gauge, the exchange-correlation energy density at a position in space is half of the electrostatic potential created there by the density of the exchange-correlation hole [21] around an electron at that position. For discussion of this and other gauges, see Refs. [22,23]. Interior scaling only makes sense if the Hartree and exchange-correlation energies are in the same gauge: If we had the exact functional, interior scaling would yield a spurious non-zero self-interaction correction to it unless the Hartree and exchange-correlation energies were in the same gauge (in which case they would cancel one another exactly).

This gauge incompatibility is present for PBE and SCAN, but not for LSDA. In the slowly-varying limit, where LDA is exact to order $\nabla^0$, the LDA energy density is uniquely defined to the same order. LDA is demonstrably in the Hartree gauge, since its energy density arises in the required way (as defined in the previous paragraph) from the exchange-correlation hole of a uniform electron gas (Eqs. (35a) and (35b) of Ref. 21). But SCAN, for example, is exact to order $\nabla^4$, while its exchange-correlation energy density is uniquely defined only to order $\nabla^0$. That is in part because the second- and fourth-order density-gradient expansions that SCAN recovers in the slowly-varying limit, if initially expressed in the Hartree gauge, have been simplified via integrations by parts.

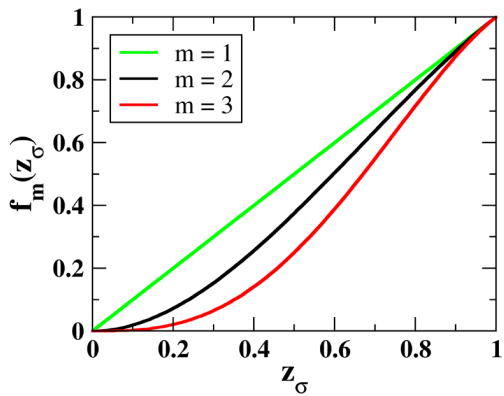

Fig. 1: The scale-down factor $f_m(z_\sigma)$ as a function of the iso-orbital indicator $z_\sigma$, applied here to PZ SIC for LDA ($m=1$), PBE ($m=2$), or SCAN ($m=3$). In this work, the same scale factor is used in sdSIC (exterior scaling) or LSIC (local or interior scaling). Note that $f = 0$ everywhere would recover the uncorrected functional, and $f = 1$ everywhere would recover PZ SIC.

An exchange-correlation energy density that is not in the Hartree gauge can be transformed to one that is by adding a compliance function $G_{\alpha\sigma}(\vec{r})$ [23,24] such that

$$\int d^3r\, G_{\alpha\sigma}(\vec{r}) = 0. \tag{9}$$

Compliance functions are also needed for the construction of local (as opposed to global) hybrid functionals [25,24]. The choice that we propose and test here is designed so that, when its approximate exchange-correlation energy is replaced by an exact one, the transformed exact exchange-correlation energy density will be in the Hartree gauge:

$$G_{\alpha\sigma}(\vec{r}) = \frac{-1}{2}n_{\alpha\sigma}(\vec{r})u([n_{\alpha\sigma}];\vec{r}) + \frac{U[n_{\alpha\sigma}]}{E_{xc}^{app}[n_{\alpha\sigma},0]}n_{\alpha\sigma}(\vec{r})\varepsilon_{xc}^{app}([n_{\alpha\sigma},0];\vec{r}), \tag{10}$$

which clearly integrates to zero. This choice guarantees that our scaled-down SIC will make no spurious non-zero correction to the exact functional. But, applied to approximate functionals, it does not perfectly achieve the Hartree gauge for one-electron densities.

Using this compliance function, the scaled-down orbital self-interaction correction can be written

$$\Delta^{sd}E_{xc}^{\alpha\sigma} =$$
$$-\frac{1}{2}\int d^3r\, f_m(z_\sigma)[n_{\alpha\sigma}(\vec{r})u([n_{\alpha\sigma}];\vec{r})]$$
$$-\int d^3r\, f_m(z_\sigma)\left[n_{\alpha\sigma}(\vec{r})\varepsilon_{xc}^{app}([n_{\alpha\sigma},0];\vec{r}) - \frac{1}{2}n_{\alpha\sigma}(\vec{r})u([n_{\alpha\sigma}];\vec{r}) + \frac{U[n_{\alpha\sigma}]}{E_{xc}^{app}[n_{\alpha\sigma},0]}n_{\alpha\sigma}(\vec{r})\varepsilon_{xc}^{app}([n_{\alpha\sigma},0];\vec{r})\right] \tag{11}$$
$$= -X_{\alpha\sigma}^{sd}\left(U[n_{\alpha\sigma}] + E_{xc}^{app}[n_{\alpha\sigma},0]\right)$$

where

$$X_{\alpha\sigma}^{sd} = \frac{\int d^3r\, f_m(z_\sigma)n_{\alpha\sigma}(\vec{r})\varepsilon_{xc}^{app}([n_{\alpha\sigma},0];\vec{r})}{\int d^3r\, n_{\alpha\sigma}(\vec{r})\varepsilon_{xc}^{app}([n_{\alpha\sigma},0];\vec{r})}. \tag{12}$$

We refer to this scaled-down SIC as sdSIC. Note that $0 \leq X_{\alpha\sigma}^{sd} \leq 1$. A remarkable outcome of the gauge transformation is that what began as an *interior*, or point-by-point, scaling of the self-interaction correction, became an *exterior* scaling. This happens because the Hartree energy $U[n_{\alpha\sigma}]$ is a fully nonlocal functional. An alternative exterior scaling scheme was proposed by Vydrov *et al.* [17], but with a *posited* scaling factor

$$X_{\alpha\sigma} = \frac{\int d^3r\, z_\sigma^k n_{\alpha\sigma}(\vec{r})}{\int d^3r\, n_{\alpha\sigma}(\vec{r})} \tag{13}$$

where $k = 1, 2$, or 3 is an adjustable parameter. In this form, increasing $k$ strongly scales down the self-interaction for all $z_\sigma$ between 0 and 1, not just for $z_\sigma \ll 1$. $X_{\alpha\sigma}$ decreases as $k$ increases, as shown in Table I of Ref. [17] for the Ar atom.

The scaled-down self-interaction correction (sdSIC) $\Delta^{sd} E_{xc}^{\alpha\sigma}$ has several correct features. First, it reduces to the full PZ-SIC correction and hence becomes exact for all one-electron densities. Second, it yields no self-interaction correction for a uniform or slowly-varying density. And finally, if $E_{xc}^{app}[n_\uparrow, n_\downarrow]$ is replaced by $E_{xc}^{exact}[n_\uparrow, n_\downarrow]$, $\Delta^{sd} E_{xc}^{\alpha\sigma}$ reduces to 0 and the correction vanishes. One incorrect feature of the sdSIC correction is that the asymptotic exchange-correlation potential seen by the outermost electron in a localized system is $-\frac{X_{HO}^{sd}}{r}$[17], where HO labels the highest occupied orbital, and not $\frac{-1}{r}$ as it should be. This occurs because our model for the compliance function $G_{\alpha\sigma}(\vec{r})$ is not semilocal, although our model for $E_{xc}^{app}[n_\uparrow, n_\downarrow]$ is.

## 3. Methodological details

To apply the DFA-sdSIC scheme self-consistently, variations of $X_{\alpha\sigma}^{sd}$ with respect to the occupied orbitals must be included in the derivation leading to Schrödinger-like SIC equations. While this is possible in principle, we expect the effect to be minor and we begin here with the following non-selfconsistent implementation to test the method. First, we carry out a standard (PZ SIC), selfconsistent DFA-SIC calculation using the Fermi-Löwdin orbital self-interaction correction (FLO-SIC) approach [26,27,28]. Then we evaluate the DFA-sdSIC and DFA-LSIC energies using the resulting Fermi-Löwdin orbitals. Those orbitals are real, although complex orbitals can to some extent mitigate the errors of PZ SIC [29,30,8]. FLO-SIC implements the PZ-SIC energy expression in a computationally efficient way. Minimizing the total energy in a FLO-SIC calculation requires solving the one-electron PZ-SIC equations self-consistently, but also optimizing a set of parameters called Fermi orbital descriptors (FODs), which are positions in three-dimensional space used in the definition of the localized Fermi-Löwdin orbitals (FLOs). Optimizing the FOD positions corresponds to finding the FLOs corresponding to the lowest FLO-SIC total energy. We then use the resulting self-consistent FLO densities $n_{\alpha\sigma}$ to compute $\Delta^{sd} E_{xc}^{\alpha\sigma}$ and the sdSIC total energy

$$E^{DFA-sdSIC} = E^{DFA} + \sum_{\alpha\sigma} \Delta^{sd} E_{xc}^{\alpha\sigma}. \qquad (14)$$

Here $E^{DFA}$ is the uncorrected total energy for a given DFA.

The results presented below were obtained using a developmental version of the FLOSIC code [31]. The default Gaussian orbital basis sets in FLOSIC [32] are of roughly quadruple-$\zeta$ quality. We used the default bases for all calculations except those for anions, where we included additional long-range functions to capture the more diffuse character of the anion orbitals. We use a selfconsistency convergence criterion of $1.0 \times 10^{-6}$ Hartree for total energy calculations, and an FOD force convergence criterion of $5.0 \times 10^{-4}$ Hartree/Bohr. All calculations were done using a very fine "variational" [33] integration grid. For calculations involving the SCAN functional, especially dense grids were used [12].

## 4. Results

In this section, we will compare several properties of atoms and molecules computed with the LDA, PBE, and SCAN functionals, without SIC and with three flavors of SIC: the exteriorly scaled-down SIC (sdSIC) of this work, the locally-scaled SIC (LSIC), and the unscaled PZ-SIC (SIC). For sdSIC and LSIC, we use the scale-down factors $f_m(z_\sigma)$ where $m$ =1, 2, or 3 for LDA, PBE, and SCAN, respectively. The LDA-LIC results are good, but many of them have been presented and discussed before [20], so they will not be further discussed before section 6. The properties are total energies of atoms, ionization energies of atoms, atomization energies of molecules, barrier heights to chemical reactions and other reaction energies that display large self-interaction errors, electron affinities of molecules, ionization energies of molecules, and equilibrium bond lengths of molecules. The data sets of our Tables I-V were also employed in Ref. [20], while those of our Tables VI (G2-1 electron affinities), VII (G2-1 ionization energies), and VIII (equilibrium bond lengths) were not.

We begin by considering the performance of sdSIC for atomic total energies. We focus here and below on the SCAN functional for illustration. Results for LDA and PBE are summarized in tables below. In Fig. 2 we present the errors of the energies for H to Ar computed in SCAN, SCAN-sdSIC, SCAN-LSIC, and SCAN-SIC. The errors are relative to the corresponding accurate total energies of Ref. [34]. For the light atoms up to carbon, all three methods reproduce the reference energies well. The SCAN-SIC energies begin to diverge from these at nitrogen (atomic number AN = 7) and significantly overestimate the reference energies for second-row atoms. The correction terms in PZ-SIC are positive for the SCAN functional, due to SCAN's too-negative exchange energies for the densities of lobed orbitals [8]. PZ-SIC changes most of the energies in the right direction, but the corrections are too large, resulting in worse agreement with the reference energies. In SCAN-sdSIC, the corrections are still positive, but much smaller, resulting in total energies that are in much better agreement with the reference values. The mean absolute errors (MAE) relative to the reference energies for SCAN (0.019 Ha), SCAN-sdSIC (0.033 Ha) and SCAN-SIC (0.147 Ha) are summarized in Table I. The values indicate that sdSIC reduces the overcorrection of SIC and results in energies that are nearly as accurate as for SCAN. Similar improvements can be seen in Table 1 for LDA and PBE as well. The LDA-sdSIC and PBE-sdSIC MAE values are 0.042 and 0.067 Ha, respectively.

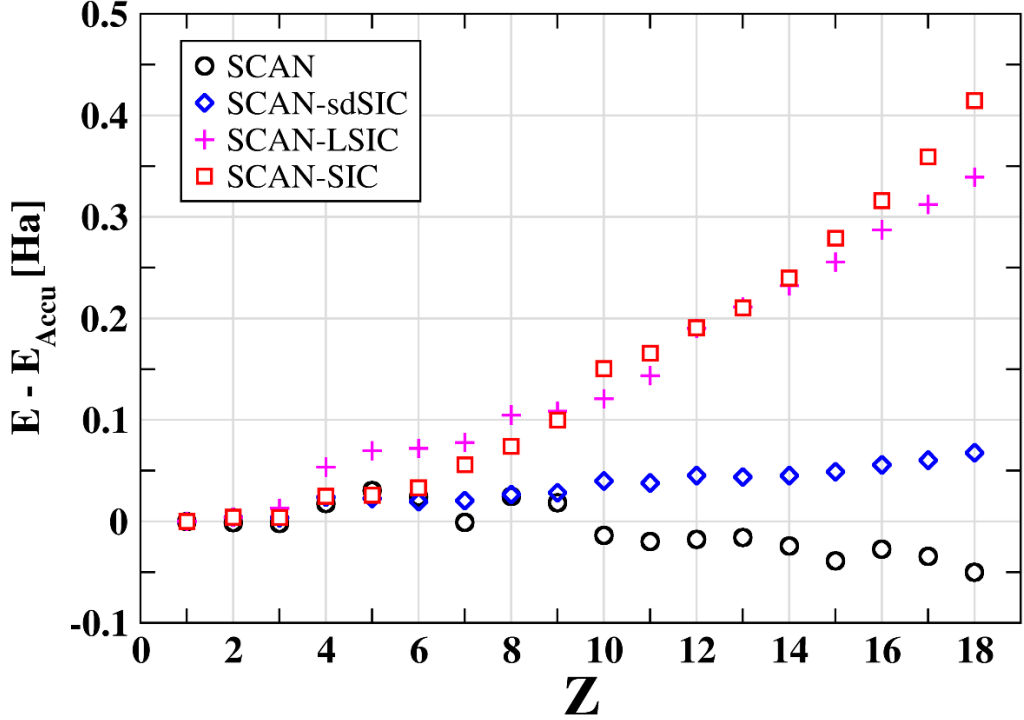

Fig. 2: Errors of total energies (in Ha) of the atoms from H to Ar, with reference energies from Ref. [34], in SCAN, SCAN-sdSIC, SCAN-LSIC, and SCAN-SIC. Z is the atomic number.

**Table I** Mean absolute errors (MAE in Ha) for total energies of the atoms from H to Ar, with reference energies from Ref. [34]. (1 Ha = 27.211 eV = 627.5 kcal/mol)

| Method | MAE (Ha) |
| --- | --- |
| LDA | 0.726 |
| LDA-sdSIC | 0.042 |
| LDA-LSIC | 0.043 |
| LDA-SIC | 0.381 |
| | |
| PBE | 0.083 |
| PBE-sdSIC | 0.067 |
| PBE-LSIC | 0.094 |
| PBE-SIC | 0.159 |
| | |
| SCAN | 0.020 |
| SCAN-sdSIC | 0.033 |
| SCAN-LSIC | 0.144 |
| SCAN-SIC | 0.147 |

Next we consider the ionization energies of the atoms from He to Kr. We compute these for each method using the ΔSCF approach, *i.e.,* we take the absolute value of the difference between the self-consistent total energies of each atom and its cation. We then compare the computed ionization energy to the experimental values of Ref. [35]. In Table II we show the MAE in eV for the various methods, first for He-Ar and then He-Kr. Comparing the two values gives a sense of how each method performs for light versus heavy atoms.

For the lighter atoms, SCAN and SCAN-sdSIC have similar MAEs (0.175 and 0.180 eV, respectively). SCAN-SIC is somewhat worse, with an MAE of 0.274 eV. SIC also worsens the performance of PBE for the light atoms and sdSIC restores it. For the heavier atoms, the MAE for all methods increases. For all atoms from He to Kr, the SCAN-sdSIC results (MAE 0.363 eV) are somewhat worse than either uncorrected SCAN (0.273 eV) or SCAN-SIC (0.259 eV). For LDA and PBE, on the other hand, sdSIC yields a lower MAE than either the uncorrected DFA or the DFA-SIC.

**Table II** Mean absolute errors (MAE in eV) of atomic ionization energies. The ionization energies are calculated using the ΔSCF method. Reference values are from Ref. [35]. The middle column shows the results for He to Ar and the right column for He to Kr.

| Method     | Z = 2 - 18 | Z = 2 - 36 |
|------------|------------|------------|
| LDA        | 0.275      | 0.458      |
| LDA-sdSIC  | 0.176      | 0.232      |
| LDA-LSIC   | 0.206      | 0.170      |
| LDA-SIC    | 0.248      | 0.364      |
|            |            |            |
| PBE        | 0.159      | 0.253      |
| PBE-sdSIC  | 0.177      | 0.226      |
| PBE-LSIC   | 0.205      | 0.363      |
| PBE-SIC    | 0.405      | 0.464      |
|            |            |            |
| SCAN       | 0.175      | 0.273      |
| SCAN-sdSIC | 0.180      | 0.363      |
| SCAN-LSIC  | 0.345      | 0.558      |
| SCAN-SIC   | 0.274      | 0.259      |

Results for the atomization energies of the molecules in the AE6 database [37] are presented in Table III. Mean error (ME), mean absolute error (MAE), and mean absolute percentage errors (MAPE) are shown relative to accurate reference values [37]. As described elsewhere [8,12], SIC strongly worsens the remarkably good performance of SCAN for atomization energies. The ME increases in magnitude from 0.3 kcal/mol for SCAN to -24.4 kcal/mol for SIC-SCAN. sdSIC restores most of the lost performance, yielding an ME of -3.3 kcal/mol. The MAE values show a similar trend, 3.0 and 26.1 kcal/mol for SCAN and SCAN-SIC, but 5.7 kcal/mol for sdSIC. sdSIC also improves the atomization energies for LDA and PBE, yielding MAE values of 25.7 and 11.7 kcal/mol, respectively for LDA-sdSIC and PBE-sdSIC.

**Table III** Mean errors (ME in kcal/mol), mean absolute errors (MAE in kcal/mol), and mean absolute percentage errors (MAPE) of the atomization energies for the molecules in the AE6 [37] database.

| Method | ME | MAE | MAPE |
|---|---|---|---|
| LDA | 75.5 | 75.5 | 16.7 |
| LDA-sdSIC | 24.5 | 25.7 | 5.6 |
| LDA-LSIC | -0.9 | 9.3 | 3.0 |
| LDA-SIC | 53.5 | 57.8 | 9.9 |
| | | | |
| PBE | 10.6 | 13.8 | 3.8 |
| PBE-sdSIC | 6.8 | 11.7 | 4.1 |
| PBE-LSIC | -16.7 | 16.7 | 4.0 |
| PBE-SIC | -15.6 | 17.8 | 5.5 |
| | | | |
| SCAN | 0.3 | 3.0 | 1.4 |
| SCAN-sdSIC | -3.3 | 5.7 | 2.4 |
| SCAN-LSIC | -47.8 | 47.8 | 10.9 |
| SCAN-SIC | -24.4 | 26.1 | 7.0 |

Previous studies have shown that SIC improves the performance of semilocal functionals for the description of reaction barriers [36]. In Table IV we summarize the results of barrier height calculations for the reactions in the BH6 database [37]. The ME for SCAN-SIC relative to accurate reference calculations is very small, -1.0 kcal/mol. For SCAN, the ME is -7.9 kcal/mol. SCAN-sdSIC yields a value midway between the two, -4.6 kcal/mol. The MAE for SCAN and SCAN-sdSIC are 7.9 and 4.6 kcal/mol, respectively, so the barriers are consistently too small in both methods. For barrier heights, the full PZ correction to SCAN is clearly important.

**Table IV** Mean errors (ME) and mean absolute errors (MAE), both in kcal/mol, for the barrier heights of the reactions in the BH6 [37] database.

| Method | ME | MAE |
|---|---|---|
| LDA | -18.1 | 18.1 |
| LDA-sdSIC | -4.1 | 4.1 |
| LDA-LSIC | 0.6 | 1.4 |
| LDA-SIC | -5.1 | 5.1 |
| | | |
| PBE | -9.6 | 9.6 |
| PBE-sdSIC | -3.7 | 4.2 |
| PBE-LSIC | 0.2 | 1.6 |
| PBE-SIC | 0.0 | 4.6 |
| | | |
| SCAN | -7.9 | 7.9 |
| SCAN-sdSIC | -4.6 | 4.6 |
| SCAN-LSIC | 4.2 | 5.1 |
| SCAN-SIC | -1.0 | 3.0 |

Self-interaction errors are expected to be especially important for the reactions in the SIE databases [37,38]. Previously, we studied the effect of the full PZ-SIC on these reactions [10]. In Table V, we compare the performance of sdSIC for these reactions to accurate reference calculations [38,39] and to our earlier results for the DFAs and DFA-SIC [20,10]. For the SIE4x4 set, which tracks the dissociation of cation dimer compounds, SCAN-SIC yields significantly better performance than SCAN (MAE = 2.2 and 17.9 kcal/mol, respectively). SCAN-sdSIC retains nearly all of this improvement, with an MAE of 3.7 kcal/mol. A similar trend in MAEs is observed for the five cationic reactions in the SIE11 database, where the SCAN-sdSIC value (5.8 kcal/mol) is significantly smaller than SCAN (10.1 kcal/mol) and nearly as small as SCAN-SIC (5.7 kcal/mol). For the six neutral reactions in SIE11, SCAN-sdSIC gives a slightly smaller MAE than SCAN-SIC (5.7 vs 6.2 kcal/mol), both better than SCAN (9.9 kcal/mol). We find results similar to these for LDA and PBE, as shown in Table V. In all cases DFA-SIC significantly improves the performance of the uncorrected DFA, and sdSIC preserves the improvement.

**Table V** Mean absolute errors (MAE), in kcal/mol, for the SIE databases [38, 39].

| Method | SIE4x4 | SIE11 | SIE11, 5 cationic | SIE11, 6 neutral |
|---|---|---|---|---|
| LDA | 27.5 | 17.8 | 22.9 | 13.4 |
| LDA-sdSIC | 5.0 | 8.9 | 11.3 | 6.9 |
| LDA-LSIC | 2.6 | 4.5 | 2.3 | 6.3 |
| LDA-SIC | 3.0 | 11.7 | 14.8 | 9.0 |
| | | | | |
| PBE | 23.3 | 11.7 | 12.7 | 10.9 |
| PBE-sdSIC | 5.9 | 7.2 | 8.9 | 5.8 |
| PBE-LSIC | 3.9 | 3.8 | 2.9 | 4.6 |
| PBE-SIC | 3.4 | 7.5 | 8.9 | 6.4 |
| | | | | |
| SCAN | 17.9 | 10.1 | 10.4 | 9.9 |
| SCAN-sdSIC | 3.7 | 5.8 | 6.0 | 5.7 |
| SCAN-LSIC | 3.8 | 11.1 | 13.5 | 9.1 |
| SCAN-SIC | 2.2 | 5.7 | 5.1 | 6.2 |

**Table VI** Mean errors (ME) and mean absolute errors (MAE), both in eV, for the electron affinities of the 7 atoms and 18 molecules in the G2-1 database, with reference values from [40].

| Method | ME | MAE |
|---|---|---|
| LDA | 0.249 | 0.255 |
| LDA-sdSIC | -0.012 | 0.146 |
| LDA-LSIC | 0.067 | 0.138 |
| LDA-SIC | -0.062 | 0.243 |
| | | |
| PBE | 0.055 | 0.087 |
| PBE-sdSIC | -0.197 | 0.199 |
| PBE-LSIC | -0.054 | 0.149 |
| PBE-SIC | -0.571 | 0.580 |
| | | |
| SCAN | -0.013 | 0.177 |
| SCAN-sdSIC | -0.120 | 0.225 |
| SCAN-LSIC | -0.059 | 0.248 |
| SCAN-SIC | -0.392 | 0.408 |

Results for the adiabatic electron affinities of the G2-1 set of 7 atoms and 18 molecules calculated with the ΔSCF method using the Sadlej basis set [41] are presented in Table VI. (This basis set includes long-range functions that better capture the extended nature of the anion orbitals.) Geometries are taken from the GMTKN55 database [39], and reference values from Curtiss *et al.* [40]. Here, because the DFAs sometimes fail to bind a full extra electron to a neutral atom or molecule, all DFA calculations are non-selfconsistent single-shot calculations using SCAN-SIC densities. Surprisingly, PBE gives the smallest MAE for these electron affinities.

**Table VII** Mean errors (ME) and mean absolute errors (MAE), in eV, for the ionization potentials of the 14 atoms and 21 molecules in the G2-1 database [40] that were carried over to the GMTKN55 database [39], with experimental reference values from [42].

| Method | ME | MAE |
| --- | --- | --- |
| LDA | 0.118 | 0.269 |
| LDA-sdSIC | 0.158 | 0.338 |
| LDA-LSIC | 0.116 | 0.307 |
| LDA-SIC | 0.303 | 0.445 |
| | | |
| PBE | -0.035 | 0.207 |
| PBE-sdSIC | 0.035 | 0.293 |
| PBE-LSIC | 0.028 | 0.213 |
| PBE-SIC | -0.204 | 0.461 |
| | | |
| SCAN | -0.038 | 0.252 |
| SCAN-sdSIC | -0.051 | 0.237 |
| SCAN-LSIC | -0.061 | 0.347 |
| SCAN-SIC | -0.233 | 0.361 |

Table VII shows the mean error (ME) and mean absolute error (MAE) in the ionization potentials, calculated using the adiabatic ΔSCF procedure, for the 14 atoms and 21 molecules in the G2-1 dataset [40] which were carried over to the GMTKN55 database [39]. We compare our numbers with the adiabatic experimental ionization potentials [42]. Surprisingly, PBE gives the smallest MAE for these ionization energies.

**Table VIII** Mean errors (ME) and mean absolute errors (MAE), in angstrom, for the equilibrium bond lengths of a benchmark set [17] of 11 diatomic molecules.

| Method | ME | MAE |
|---|---|---|
| LDA | 0.0076 | 0.0110 |
| LDA-sdSIC | -0.0085 | 0.0189 |
| LDA-LSIC | -0.0015 | 0.0129 |
| LDA-SIC | -0.0317 | 0.0392 |
| | | |
| PBE | 0.0123 | 0.0123 |
| PBE-sdSIC | -0.0019 | 0.0132 |
| PBE-LSIC | 0.0015 | 0.0114 |
| PBE-SIC | -0.0134 | 0.0257 |
| | | |
| SCAN | 0.0039 | 0.0057 |
| SCAN-sdSIC | -0.0110 | 0.0110 |
| SCAN-LSIC | -0.0010 | 0.0146 |
| SCAN-SIC | -0.0190 | 0.0197 |

Table VIII shows the mean error and mean absolute error of the equilibrium bond lengths for the benchmark set [17] of 11 diatomic molecules, computed by minimization of the total energy under variation of bond length. The molecules in this set are BeH, BH, CH4, $C_2$ ($^1\Sigma^+_g$), CO, $N_2$, OH, $O_2$, HF, and $F_2$. Reference values are taken from Vydrov et. al. [17]. It can be seen that SCAN outperforms all other methods and the PZ SIC to the density functionals strongly underestimates the bond lengths. However, sdSIC brings down the error to a value closer to that of the parent functionals.

## 5. Discussion: sdSIC versus other scaling methods

Other research works have recognized the utility of scaling down the PZ SIC [42,17,18,30,20]. Among these, the exterior scaling of Vydrov *et al.* [17] is the most similar to the sdSIC method we introduce here. In that work, exterior scaling was chosen largely for computational ease, but it was also justified as a way to avoid the potential problem of gauge-inconsistency in the energy densities of the self-Hartree and self-exchange-correlation terms. Like sdSIC, the method of Ref. [17] has an external scaling factor for each orbital, but the scaling factor (Eq. (13)) is *posited* (not derived) in Ref. [17]. Here we *derive* a different expression (Eq. (12)). As discussed in section 2, our new scale-down factor $f_m(z_\sigma)$ (Eq. (8)) is better suited than the $z_\sigma^k$ used in Ref. [17] for drawing conclusions about the importance of the slowly-varying limit to atoms and molecules. In Ref. [17], the value of *k* was varied to obtain the best agreement between the scaled PZ-SIC results and reference values. For example, for the total energies of the atoms from Li to Ar, the smallest MAEs were obtained using *k* = 1 and 3 for LDA and PBE, respectively. On the other hand, for the atomization energies of the AE6 set, *k* =3 and ½ gave the smallest MAEs for LDA and PBE respectively. Our motivation in developing sdSIC is somewhat different. sdSIC is derived, via a gauge transformation that ensures no correction to the exact functional, and its scale-down factor is chosen to keep the description of only the slowly-varying limit $z_\sigma \ll 1$ the same in DFA-sdSIC as in DFA. The improvements that DFA-sdSIC yields over DFA-SIC, in particular those shown in Tables I-III above, can then be attributed to the correction of errors in the slowly-varying limit caused by DFA-SIC. This success can be taken as support for a continued strategy of satisfying the known mathematical properties of $E_{xc}^{exact}$ in the quest for a single density functional-based method that yields accurate predictions in all settings.

## 6. Conclusions

In this paper we introduced a scaled-down self-interaction correction that uses an iso-orbital indicator to scale down the Perdew-Zunger SIC in regions where the density is slowly-varying, but leaves it intact in regions where the density is one-electron like. We showed that making this correction gauge-consistent for the exact functional leads from an interior or local scaling to an external scaling of the SIC terms. The resulting scaled-down SIC (sdSIC) method is exactly self-interaction free for a one-electron density, and gives zero correction in the limit of a slowly-varying density, where the LDA, PBE, and SCAN functionals are already correct over a range that increases from LDA to PBE to SCAN. The results presented above show that sdSIC, used in conjunction with the highly accurate SCAN functional (*i.e.,* SCAN-sdSIC), improves the performance of SCAN in situations where self-interaction errors are important, although not quite as well as full SCAN-SIC, and restores much of the accuracy of the SCAN description of molecular atomization energies that is severely degraded in SCAN-SIC. The exteriorly-scaled SCAN-sdSIC performs rather well. Similar behavior is seen for LDA-sdSIC and PBE-sdSIC.

Taken together, our results show that that the largest errors of DFA-SIC for the equilibrium properties of molecules can be corrected by restoring the correct slowly-varying limits of the uncorrected DFAs. This confirms the relevance [19] of the slowly-varying-in-space limit to real atoms and molecules.

Finally, we have demonstrated that local-scaling SIC (LSIC) works remarkably well as a correction to LDA, and not so well as a correction to PBE and especially SCAN, and that this is so because the exchange-correlation energy density is in the Hartree gauge for LDA but not for PBE or SCAN. Building a good local-scaling self-interaction correction to PBE or SCAN requires the development of good semilocal compliance functions $G_{\alpha\sigma}$ for them.

LDA-LSIC is typically better than LDA-sdSIC, demonstrating once again that the Hartree gauge for the exchange-correlation energy is important for local-scaling approaches. sdSIC has been introduced here primarily to improve our understanding of density functional approximations. LDA-LSIC is conceptually simpler and often more accurate for total energy differences. The Appendix of Ref. [17] would make it practical to implement sdSIC self-consistently, while the self-consistent implementation of LSIC may be more challenging.

Although SCAN is found here to be the best functional for equilibrium properties, a remarkably good and well-balanced performance for the equilibrium and stretched-bond properties of atoms and molecules seems to be achieved by LDA-LSIC. LDA-LSIC satisfies only two exact constraints (plus others that those two imply): the uniform-gas and one-electron limits, suggesting that these may be the most important exact constraints. In more precise language [3], these two limits may be the most-appropriate and nearly-sufficient norms for a nonlocal functional. The other exact constraints built into the semilocal functionals like PBE and SCAN are in a sense building up an approximate self-interaction correction to LDA that can be excellent for equilibrium properties but not so good for stretched-bond properties.

While not as good overall as LDA-LSIC, PBE-LSIC is not bad, and is almost always better than PBE-SIC. It may be that the gauge-inconsistency error in PBE-LSIC is less problematic than the error of PBE-SIC for slowly-varying density.

All self-interaction corrected results here have been obtained with real (noded for systems with $z_\sigma < 1$) localized SIC orbitals. Some further improvement in these results might be found with complex (lobed but un-noded) orbitals [29,30,8]. Complex orbitals are not available in the current version of the FLOSIC code [31].

**Acknowledgments:** This work was supported by the U.S. Department of Energy, Office of Science, Office of Basic Energy Sciences, as part of the Computational Chemical Sciences Program under Award No. DE-SC0018331. The work of PB, KW, and CS was supported by the U.S. National Science Foundation under Grant No, DMR-1607868 to JPP (CMMT – Division of Materials Theory, with a contribution from CTMC – Division of Chemistry). This research includes calculations carried out on HPC resources supported in part by the National Science Foundation through major research instrumentation grant number 1625061 and by the US Army Research Laboratory under contract number W911NF-16-2-0189.

The data that supports the findings of the study are available from the corresponding author upon reasonable request.